\documentclass[]{jpsj3}
\usepackage{comment}
\usepackage{txfonts}
\usepackage{braket,ulem}
\usepackage[dvipdfmx]{color}
\usepackage{booktabs}
\title{Possible Excitonic Phase of Graphite in the Quantum Limit State}

\author{Kazuto Akiba\thanks{k{\_}akiba@issp.u-tokyo.ac.jp}, Atsushi Miyake, Hiroshi Yaguchi$^1$, Akira Matsuo, 
Koichi Kindo, and Masashi Tokunaga}
\inst{The Institute for Solid State Physics, The University of Tokyo, Kashiwa, Chiba 277-8581, Japan \\
$^1$Department of Physics, Faculty of Science and Technology, Tokyo University of Science, Noda, Chiba 278-8510, Japan \\
} %\\

\abst{The in-plane resistivity, Hall resistivity and magnetization of graphite were investigated in pulsed magnetic fields applied along the \textit{c}-axis. The Hall resistivity approaches zero at around 53 T where the in-plane and out-of-plane resistivities steeply decrease. The differential magnetization also shows an anomaly at around this field with a similar amplitude compared to that of de Haas-van Alphen oscillations at lower fields. 
This transition field appears insensitive to disorder, but reduces with doping holes.
These results suggest the realization of the quantum limit states above 53 T. 
As a plausible explanation for the observed gapped out-of-plane conduction above 53 T, the emergence of the excitonic BCS-like state in graphite is proposed.
}

\kword{graphite, high magnetic fields, density wave, magneto-transport, magnetization}
\begin{document}
\maketitle

\section{Introduction}
Graphite consists of stacked graphene layers along the \textit{c}-axis, and is known as a compensated semimetal having almost the same number of electron and hole carriers. The carrier densities are about $3 \times 10^{18}$ cm$^{-3}$ for both carriers, which is considerably smaller than that in typical metals. Elongated Fermi surfaces are located along the H-K-H (H'-K'-H') points in the six edges of the hexagonal Brillouin zone. The electron-like Fermi surfaces exist near the K (K') points and the hole-like ones near the H (H') points. The band structure along the H-K-H (H'-K'-H') points is well explained by the Slonczewski-Weiss-McClure (SWM) model. \cite{SW,Mc}

In magnetic fields applied parallel to the \textit{c}-axis, the band structure is quantized into Landau sub-bands, which show dispersion only along the \textit{c}-axis. Owing to the small effective mass of carriers in the basal plane, the quasi-quantum limit state, where only the lowest electron-like (Landau index of $n=0$) and hole-like ($n=-1$) Landau sub-bands (each of them spin-split) cross the Fermi level, is realized in a moderate magnetic field of about 7.4 T. 

\begin{figure}
\begin{center}
\includegraphics[width=7.5cm]{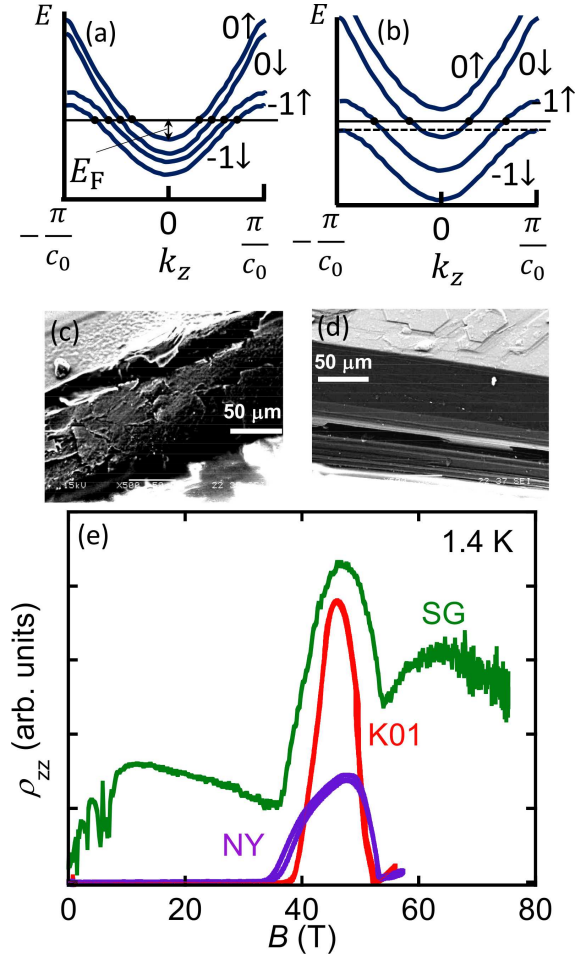}
\caption{(Color online) 
Schematic drawings of the band structures (a) below and (b) above 53 T calculated by TG.\cite{TG} The horizontal solid lines indicate the Fermi level of pristine graphite while the dotted line in (b) indicates the Fermi level of the hole-doped one.
SEM pictures of the sample edges (c) cut by razor and (d) as-grown crystal of K01 samples.
(e) The out-of-plane magnetoresistance $\rho_{zz}$ of as-grown K01, natural NY, and SG samples.
}
\label{picandzz}
\end{center}
\end{figure}

The application of higher magnetic fields to the quasi-quantum limit state in graphite causes an abrupt increase in the in-plane resistance above $\sim$ 25 T.\cite{Tanuma} Since the onset field strongly depends on temperature, this anomaly is ascribed to an electronic phase transition involving many-body effects.\cite{manybody} A theoretical study by Yoshioka and Fukuyama (YF) suggested that the formation of a charge-density-wave state caused by Fermi-surface nesting in the $n=0$ spin-up (0$\uparrow$) sub-band, and reasonably reproduced the temperature dependence of the transition field.\cite{YF} 
Here, the direct Coulomb interaction was eliminated by assuming the anti-phase charge modulation along the H-K-H and H'-K'-H' edges. 
According to this model, the transition temperature for the density-wave state $T_\mathrm{c}$ is given in the mean-field approximation as
\begin{equation}
k_\mathrm{B} T_\mathrm{c}=4.53\,E_\mathrm{F} \,\frac{\cos^2 (c_0k_{\mathrm{F0\uparrow}}/2)}{\cos(c_0 k_{\mathrm{F0\uparrow}})}\,\mathrm{exp}\left( -\frac{2}{N_{0\uparrow}(E_\mathrm{F})u(\epsilon)}\right),\label{YFformula}
\end{equation}
where $E_\mathrm{F}$ is the Fermi energy measured from the bottom of the Landau sub-bands as shown in Fig. \ref{picandzz}(a) and the $N_{0\uparrow}(E_\mathrm{F})$ is the density of states at the $E_\mathrm{F}$ for the $0$$\uparrow$ sub-band. $k_{\mathrm{F0\uparrow}}$ is the Fermi wavevector for this sub-band and $c_0$ is the lattice constant along the \textit{c}-axis. $u(\epsilon)$ represents the relevant pairing interaction as a function of the dielectric constant ($\epsilon$). In this formula, all the parameters $E_\mathrm{F}$, $k_{\mathrm{F0\uparrow}}$, $N_{0\uparrow}(E_\mathrm{F})$, and $u(\epsilon)$ depend on magnetic field.
On the other hand, Takahashi and Takada claimed that the spin-density-wave state, characterized by another nesting vector, becomes stable when the spatial separation of electrons along the H-K-H and H'-K'-H edges is taken into account.\cite{TT} Since there is no report on the direct experimental evidence of the nesting vector, the real nature in this field region remains as an open question.

While Eq. (\ref{YFformula}) is given for a particular 2$k_\mathrm{F}$-instability, we remark that Eq. (\ref{YFformula}) has a similar functional form to the general BCS-type formula for mean-field-type pairing\cite{Yaguchireview}
\begin{equation}
k_\mathrm{B}T_\mathrm{c}=1.13\,E_\mathrm{F}\,\mathrm{exp} \left(-\frac{1}{N(E_\mathrm{F})V}\right), \label{BCS}
\end{equation}
where $N(E_\mathrm{F})$ is the density of states at the Fermi level and $V$ is the relevant pairing interaction.
When the Fermi energy approaches zero, $T_\mathrm{c}$ goes to zero in Eq. (\ref{BCS}).
Consequently, a density-wave state with $T_\mathrm{c}$ expressed by the Eq. (\ref{BCS}) is expected to exhibit a reentrant transition back to the normal state if $E_\mathrm{F}$ of the relevant sub-band approaches zero in higher magnetic fields. Takada and Goto calculated the renormalized band structure taking into account self-energy effects, and found that the $0$$\uparrow$ and $-1$$\downarrow$ Landau subbands detouch the Fermi level almost simultaneously at a crossing field of $\sim$ 53 T\cite{TG} (see Fig. \ref{picandzz}(b)).

In fact, measurements of the in-plane resistance in magnetic fields of up to $\sim$ 55 T obtained a clear indication of such a reentrant transition; the in-plane and out-of-plane resistance shows a sharp bend at a field close to the crossing field of $\sim$ 53 T.\cite{Yaguchi,Yaguchi_pb}

\begin{comment}
The transition fields to the density wave state and reentrant transition fields to the normal state are sensitively affected by carrier imbalance and/or lattice defects.
Neutron irradiation to graphite introduces 1) a pair breaking effect and 2) hole doping effect simultaneously by creating lattice defects.
Yaguchi \textit{et al.} investigated the magneto-resistance of neutron-irradiated Kish graphite and found the temperature dependence of the transition field deviates from the YF theory as irradiation times becomes long.\cite{Neutron} They pointed out that this behavior is due to interruption of electron-electron (CDW) or electron-hole (SDW) pairing by lattice defects. It was also reported that the re-entrant transition field are significantly lower than that of pristine graphite.\cite{Neutron_re} 
This fact indicates that the electron sub-band is responsible for the formation of density wave state. 
\end{comment}

Recently Fauqu{\'e} \textit{et al}. investigated the longitudinal magneto-resistance ($\rho_{zz}$) in fields up to 80 T, and found an additional enhancement of the $\rho_{zz}$ above 53 T and successive reduction of it above 75 T.\cite{Fauque} 
The authors interpreted these anomalies as sequential depopulation of Landau subbands, depopulation of hole-like $-1$$\downarrow$ subband at 53 T and electron-like $0$$\uparrow$ one at 75 T, and the emergence of novel density wave state in the intermediate field contrary to the preceding scenario. In this field region, the out-of-plane conduction exhibits gapped behavior, while the in-plane one ($\rho_{xx}$) remains metallic. This anomalous anisotropic conduction was explained with assuming parallel conduction through the chiral edge state, which is characteristic of the quantum Hall state.

To elucidate the true nature of the electronic states of graphite, studies of the Hall resistance in high magnetic fields will be important. In highly oriented pyrolytic graphite (HOPG), Kopelevich \textit{et al.} observed multiple anomalies in Hall resistivity in field region between 16 T and 43 T, and ascribed them to fractional quantum Hall plateaus.\cite{Kopelevich}
They also reported the sign reversal in Hall resistance at 43 T, and interpreted as the change of majority carrier from electron to hole.
Later, Kumar \textit{et al.} reported that similar sign reversal occurs even at room temperature,\cite{Kumar} suggesting the irrelevance of the sign reversal of Hall resistance and the density wave state in HOPG.
In experiments on single crystalline Kish graphite up to 30 T, the absolute value of Hall resistance steeply decreases in the density wave state.\cite{Uji} The result does not show any feature of fractional Hall plateaus at least up to 30 T. The Hall resistance of single crystalline graphite above 30 T has not been reported as yet.

To obtain further understanding of the true electronic states of graphite in the field region over 30 T, we carried out simultaneous measurements of the magneto-resistance and Hall resistance up to 65 T and of the magnetization up to 75 T.
\section{Experiment}

In this study, we investigated the transport and magnetic properties in various types of graphite in high magnetic fields applied along the \textit{c}-axis. 
We investigated three batches of Kish graphite (K01, K02, and K03), one super graphite (SG) made by a graphitization of piled polyimide sheets, and one natural New York graphite (NY).
%All investigated samples are shown in Table \ref{tab_allsample}.

Since  the in-plane chemical bonding is significantly stronger than the out-of-plane coupling, mechanically shaped graphite usually has disordered edges as shown in Fig. 1(c). Therefore, we mainly used as-grown crystals having clean edges as shown in Fig. 1(d) to extract intrinsic transport properties.
In addition, graphite is known as a highly anisotropic conductor showing non-linear $I-V$ characteristics in the field-induced density-wave state.\cite{NL_1,NL_2}
To avoid possible mixing of in-plane and out-of-plane conductions,\cite{Qing,Levin} we used thin samples with the ratio of length/thickness $\sim$ 50 - 100 and formed electric contacts on the top surfaces of them for in-plane transport measurements. In this configuration, possible partial cleavage in the sample might affect in the evaluation of the absolute values of $\rho_{xx}$ and $\rho_{xy}$. Thereby, we do not discuss quantitative aspect of the resistivity in this paper.

Figure 1(e) shows the out-of-plane magneto-resistance ($\rho_{zz}$) of K01, NY, and SG.
All the samples show additional increase in $\rho_{zz}$ above 53 T as reported previously.\cite{Yaguchi,Fauque}
Therefore, we regard this enhancement as intrinsic nature in graphite.

In-plane ($\rho_{xx}$) and Hall resistance ($\rho_{xy}$) of the samples were measured simultaneously by a five-probe method using the numerical lock-in technique\cite{Machel} at a frequency of 100 kHz or 200 kHz. Epo-tek H20E silver epoxy and  30 $\mu$m-thick gold wires were used to form electrical contacts. 
Magnetization ($M$) was measured by an induction method using two pickup coils placed coaxially. To achieve large signal-to-noise ratio in the magnetization measurement, the crystals of K01 were piled up with their \textit{c}-axis aligned as shown in the inset of Fig. \ref{ManddMdB_re}(a).
Magnetic fields were applied along the \textit{c}-axis throughout the whole measurements.

Pulsed high magnetic fields were generated using non-destructive-type magnets installed at The Institute for Solid State Physics at The University of Tokyo. For magnetotransport measurements, magnetic fields up to 56 T and 65 T (pulse duration of 36 ms and 4 ms, respectively) were generated using bipolar pulse magnets, which can generate both positive and negative fields. Mixing of the $\rho_{xx}$ component into the nominal $\rho_{xy}$ signal was eliminated by subtracting the data for different field polarities. Magnetization and some of $\rho_{zz}$ were measured up to 75 T using a unipolar pulse magnet with a duration of 4 ms.
 
\section{Results and discussion}
\begin{figure}
\begin{center}
\includegraphics[width=7.5cm]{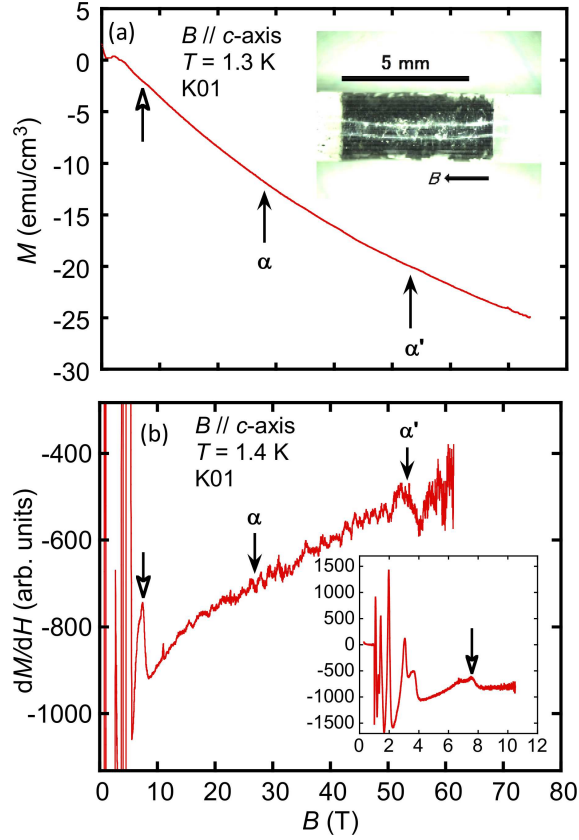}
\caption{(Color online) (a) The field dependence of magnetization of K01 sample up to 74 T at 1.3 K. The inset shows a picture of the sample used for the magnetization measurements.
(b) The field dependence of differential magnetization of K01 up to 62 T. $\alpha$ and $\alpha'$ indicate the fields where in-plane magneto-resistivity shows anomalies, while the open arrow indicates the field at which graphite goes into quasi quantum limit state. The inset shows the differential magnetization in the low field region.}
\label{ManddMdB_re}
\end{center}
\end{figure}

Figure \ref{ManddMdB_re}(a) shows the field dependence of magnetization ($M$) of K01. Monotonic sub-linear diamagnetism is observed up to 74 T.
Figure \ref{ManddMdB_re}(b) shows the differential magnetization d$M$/d$H$ obtained by averaging seven data sets acquired under repeated experiments up to 62 T. The d$M$/d$H$ curve up to 11 T is shown in the inset of Fig. \ref{ManddMdB_re}(b). Oscillations in d$M$/d$H$ observed in the region up to 7.4 T corresponds to the de Haas-van Alphen (dHvA) effect indicating all sub-bands except for $n=0$ and $-1$ detach from the Fermi level at fields below 7.4 T. In the main panel, we can hardly identify anomaly at the transition field $\alpha$ shown below. On the other hand, a non-monotonic change was observed at around 53 T. 

\begin{figure}
\begin{center}
\includegraphics[width=7.5cm]{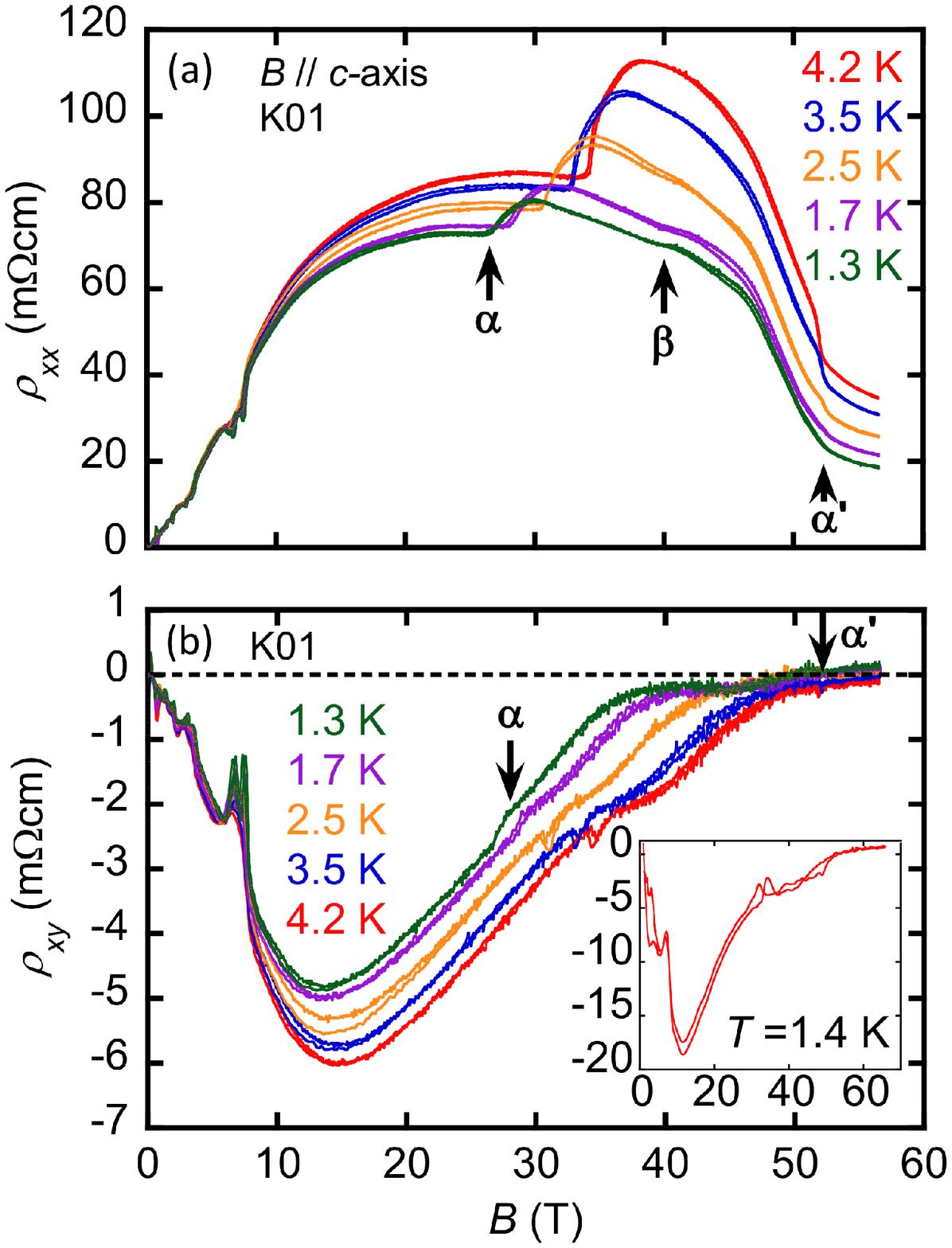}
\caption{(Color online) (a) In-plane resistivity and (b) Hall resistivity of K01 sample at various temperatures as functions of magnetic fields applied along the \textit{c}-axis. The inset of (b) shows the Hall resistivity up to 65 T at 1.4 K. The result shown in inset was obtained for a different piece of K01 from that in the main panel.}
\label{xxandxy}
\end{center}
\end{figure}

Figure \ref{xxandxy}(a) shows the in-plane magneto-resistance $\rho_{xx}$ of K01 at various temperatures between 1.3 K and 4.2 K. Shubnikov-de Haas oscillations are observed in magnetic fields below 7.4 T. In higher fields, the observed abrupt increase and steep decrease in $\rho_{xx}$ (marked with $\alpha$ and $\alpha'$) have been regarded as the emergence and collapse of the density wave state, respectively. The critical field of the $\alpha$ transition shows strong temperature dependence indicating the importance of the many-body effect. On the other hand, the critical field of the $\alpha'$ transition show rather weak temperature dependence as reported previously.\cite{Yaguchi} 
Meanwhile, the other anomaly seen between $\alpha$ and $\alpha'$ pointed out in the previous reports\cite{Yaguchi,Ochimizu} (which is referred to as $\beta$ anomaly below) is quite small in $\rho_{xx}$. We observed clear $\beta$ anomaly for the relatively thick samples (not shown). This geometrical effect implies that the $\beta$ anomaly might be the superposition of the out-of-plane resistivity. We will not go into details about the $\beta$ anomaly in this study. The $\rho_{xx}$ above 53 T decreases with decreasing the temperature contrary to the gapped behavior in $\rho_{zz}$, as reported previously.\cite{Fauque}

Figure \ref{xxandxy}(b) shows the field dependence of Hall resistivity $\rho_{xy}$ at temperatures between 1.3 K and 4.2 K. Shubnikov-de Haas oscillations are also observed in the $\rho_{xy}$ below 7.4 T. In higher fields, an anomaly is clearly observed at $\alpha$, but is less pronounced than that in $\rho_{xx}$. 
Some readers might concern the relevance of this anomaly with the quantum Hall plateau. 
Since the normalized Hall resistance per single graphite layer at $\alpha$ varies from sample to sample in the other sets of experiments on K01, we consider that this plateau-like anomaly at around $\rho_{xy}$ $\sim$ $-2$ m$\Omega$ cm is not related to the quantum Hall plateau. Fractional plateau-like structures, which have been reported on HOPG,\cite{Kopelevich} were not observed in this measurement.
We also observed that $\rho_{xy}$ gradually goes to zero as the applied field approached 53 T. As shown in the inset of Fig. \ref{xxandxy}(b), $\rho_{xy}$ remains almost zero even deeply in the high field state between 53 T and 74 T.  

\begin{figure}
\begin{center}
\includegraphics[width=7.5cm]{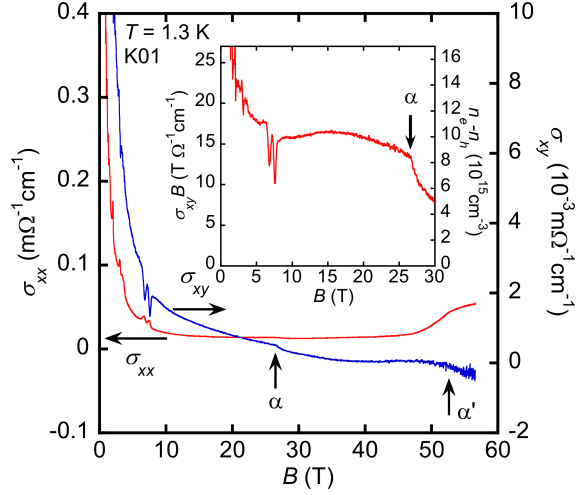}
\caption{(Color online) The field dependence of conductivities $\sigma_{xx}$ (red) and $\sigma_{xy}$ (blue) of K01 sample at 1.3 K. The inset shows the field dependence of $\sigma_{xy}B$ up to 30 T at 1.3 K. The scale on the right side of inset gives a difference between electron- and hole-density. }
\label{sigma}
\end{center}
\end{figure}

\begin{figure}
\begin{center}
\includegraphics[width=15cm]{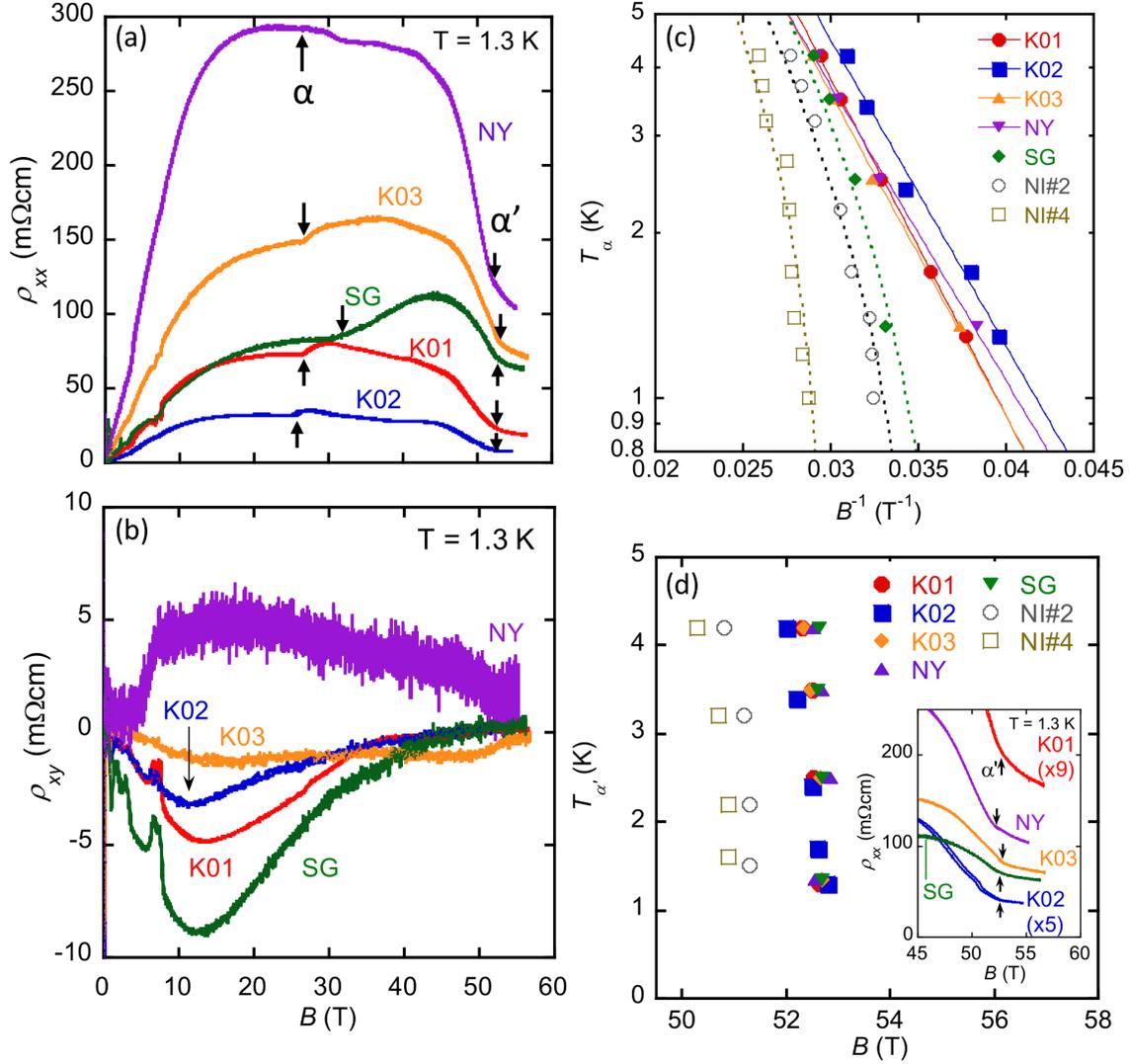}
\caption{
(Color online) (a) The in-plane magneto-resistance $\rho_{xx}$ and (b) Hall resistance $\rho_{xy}$ of five different samples at 1.3 K. (c) The field dependence of $\alpha$ transition temperature $T_\mathrm{\alpha}$ as a function of inversed field. The open circles and squares represent the reported results of Kish graphite irradiated with fast neutron for two (NI$\#$2) and four (NI$\#$4) hours, respectively\cite{Neutron}. (d) The field dependence of $\alpha'$ transition temperature $T_{\alpha'}$ of five different pristine and two irradiated\cite{Neutron_re} graphite samples. 
NI$\#$2 and NI$\#$4 are irradiated with first neutron under the same condition, but 
different pieces from those in (c). 
The inset shows enlarged view of (a) at around the $\alpha'$ transition field.
}
\label{YFplot}
\end{center}
\end{figure}

Figure \ref{sigma} shows the field dependence of the conductivities $\sigma_{xx}$ and $\sigma_{xy}$ of K01.  The $\sigma_{xx}$ and $\sigma_{xy}$ were obtained from the in-plane resistivity $\rho_{xx}$ and Hall resistivity $\rho_{xy}$ using the following relationship:
\begin{align}
\sigma_{xx}&=\frac{\rho_{xx}}{\rho_{xx}^2 + \rho_{xy}^2} \\
\sigma_{xy}&=-\frac{\rho_{xy}}{\rho_{xx}^2 + \rho_{xy}^2}.
\end{align}
In high magnetic fields, the imbalance between the carrier densities of electron ($n_e$) and hole ($n_h$) is deduced within the framework of semi-classical simple two carrier model as follows:
\begin{equation}
n_e-n_h=\frac{\sigma_{xy}B}{e}.
\label{imbalance}
\end{equation}
The field dependence of $\sigma_{xy}B$ and thus estimated imbalance are shown in the inset of Fig. \ref{sigma}. 
The $\sigma_{xy}B$ is nearly constant in the field region between 8 T and 15 T, then shows a slight decrease between 15 T and 25 T. This reduction has been ascribed to the effect of magnetic freeze-out, which diminishes carriers provided by ionized impurities.\cite{Sugihara} 
As shown in the inset of Fig. \ref{sigma}, $\sigma_{xy}B$ shows a dramatic decrease at $\alpha$. As mentioned above, Uji \textit{et al.}\cite{Uji} ascribed this anomaly to the opening of a gap in the electron-like sub-band, which leads to the sign reversal of the Hall resistivity at higher fields. Contrary to this expectation, no apparent sign reversal was observed in our measurements up to 65 T.
To clarify whether these features are intrinsic in graphite, we measured $\rho_{xx}$ and $\rho_{xy}$ in various types of graphite [Figs. \ref{YFplot}(a), (b)]. In $\rho_{xx}$, only the $\alpha$ transition field of SG is clearly higher than the other samples. The details of this difference will be discussed later.
The $\rho_{xy}$ of K01, K02, K03, and SG are negative, whereas only natural NY shows positive Hall resistance.
Regardless of  its sign, the $\rho_{xy}$ approaches zero at the $\alpha'$ transition in all the samples, suggesting it is intrinsic behavior in graphite. No quantum-Hall plateau-like structure was observed in any samples.

\begin{table}
\caption{
The list of samples investigated in this report. $T^*$, $B^*$, and $\tau$ are fitting parameters in Eqs. (\ref{eq_emp}) and (\ref{eq_pair}).
}
\label{tab_allsample}
\begin{center}
\begin{tabular}{lrrrrr}\toprule
sample & measured quantity &$T^*$ (K)&$B^*$ (T)&$\tau$ ($\times$ 10$^{-12}$ s) &paper\\ \midrule
K01 (Kish graphite) & $\rho_{xx}$, $\rho_{xy}$, $\rho_{zz}$, $M$ & 290&144&- &this work\\
K02 (Kish graphite) & $\rho_{xx}$, $\rho_{xy}$ &270&136&-&this work\\
K03 (Kish graphite) & $\rho_{xx}$, $\rho_{xy}$ & 257&142&- &this work\\
NY (natural graphite) & $\rho_{xx}$, $\rho_{xy}$, $\rho_{zz}$ & 192&131&- &this work\\ 
SG (super graphite)& $\rho_{xx}$, $\rho_{xy}$, $\rho_{zz}$ & 270&136&4.5 &this work\\ \midrule
NI$\#$2 (Kish graphite) & $\rho_{xx}$ &191&128&3.9&Ref. \ref{ref_Neutron} and \ref{ref_Neutron_re}\\
NI$\#$4 (Kish graphite) & $\rho_{xx}$ &191&128&2.0&Ref. \ref{ref_Neutron} and \ref{ref_Neutron_re}\\
\bottomrule
\end{tabular}
\end{center}
\end{table}
Figure \ref{YFplot}(c) shows the field dependence of the transition temperature for $\alpha$ ($T_{\alpha}$). According to the BCS-like formula in Eq. (\ref{BCS}), and assuming the enhancement of the $N(E_{\mathrm{F}})$ in proportion to magnetic field and ignoring field dependence of $E_\mathrm{F}$ and $V$, Eq. (\ref{BCS}) is rewritten in simpler form: 
\begin{equation}
T_{\alpha}=T^*\exp\left(-\dfrac{B^*}{B}\right),
\label{eq_emp}
\end{equation}
where $T^*$ and $B^*$ are constants\cite{manybody}. This formula gives linear relation between $\ln\,T_\alpha$ and the inversed field.
Solid lines in Fig. \ref{YFplot}(c) represent fitting curves for K01, K02, K03, and NY by Eq. (\ref{eq_emp}) with using parameters of $T^*$ and $B^*$ as listed in Table \ref{tab_allsample}.
Here, the $T^*$ and $B^*$ simply modify the scales of vertical and horizontal axes, respectively.
In our samples, only the SG exhibits non-linear phase boundary that cannot be pronounced by Eq. (\ref{eq_emp}). Similar discrepancy was also observed in neutron-irradiated graphite\cite{Neutron}. 
Neutron irradiation to graphite is known to introduce lattice defects that act as acceptors, and hence, dope hole carriers.
The reported field dependence of $T_\alpha$ of neutron-irradiated Kish graphite are also shown in Fig. \ref{YFplot}(c)\cite{Neutron}. 
The results for irradiated samples significantly deviate from that of Eq. (\ref{eq_emp}) similar to the cases of SG.
Yaguchi \textit{et al.} explained the qualitative difference of the transition temperature with introducing the pair breaking effect on density waves caused by charged impurities\cite{Neutron}, similar to the arguments on 
different types of graphite below 1 K\cite{Iye}. In the mean-field form, the reduced transition temperature 
by the pair breaking effect ($T_\alpha^{\mathrm{PB}}$) is expressed as,\cite{Iye}
\begin{equation}
\ln \left( \dfrac{T_\alpha^{\mathrm{PB}}}{T_{\alpha}}\right)=\psi\left(\dfrac{1}{2}\right)-\psi\left(\dfrac{1}{2}+\dfrac{\hbar}{2\pi k_\mathrm{B} T_\alpha^{\mathrm{PB}} \tau }\right).
\label{eq_pair}
\end{equation} 
$\psi(x)$ is the digamma function, $T_{\alpha}$ is the transition temperature without the pair breaking effect, which is given by Eq. (\ref{eq_emp}), and $\tau$ is  the relaxation time of pair breaking.
The deviation from the functional form of Eq. (\ref{eq_emp}) seen in neutron-irradiated Kish graphite is well reproduced by Eq. (\ref{eq_pair}) as shown in Fig. \ref{YFplot}(c). Apparently Eq. (\ref{eq_pair}) also accounts for the reduction of $T_\alpha$ seen in SG. With using the values of $T^*$ = 270 K and $B^*$ = 136 T taking from those in the K02, the phase boundary of SG can be reasonably reproduced with $\tau$ = 4.5 $\times$ 10$^{-12}$ s as shown by the dotted line in Fig. \ref{YFplot}(c). Although the microscopic origin is not clear, the $T_\alpha$ in the SG is significantly affected by the pair breaking effect compared with the other four samples.

Although the $T_\alpha$ of SG shows the similar trace with those of neutron-irradiated samples, the transition temperature for $\alpha$' ($T_{\alpha'}$) shows up significantly different.
The $T_{\alpha'}$ for all samples are plotted as a function of $B$ in Fig. \ref{YFplot}(d). The five pristine samples show almost the similar boundary, whereas the neutron-irradiated ones, quoted from literature\cite{Neutron_re}, show transitions at smaller fields. We consider that the reduction of the transition field at $\alpha'$ at a given temperature is not caused by the pair breaking effect because the observed reduction of that of SG is almost the same value as pristine Kish samples. The neutron-irradiated samples are known to have holes with the order of 10$^{18}$ cm$^{-3}$, which is hundred times lager than the change imbalance for the pristine samples. Therefore, the reduction of the re-entrant transition field at $\alpha'$ in irradiated samples can be caused by the hole-doping rather than the effect of pair breaking.   

Let us discuss the transition at $\alpha'$ in the following. Firstly, since the differential magnetization shows anomaly at around 53 T in the same order of magnitude with those in the dHvA oscillations at low fields (open arrows in Fig. \ref{ManddMdB_re}).
We regard this anomaly as the feature of depopulation of the Landau sub-band. Within the accuracy of the these measurements, however, we cannot determine the number of the depopulated sub-bands at $\sim$ 53 T.

To consider the number of populated Landau sub-bands above 53 T, we discuss the result of Hall resistance beyond the simple two-carrier model. For more precise argument, we have to calculate the in-plane and Hall conductivity using the Kubo formula:
\begin{equation}
\sigma_{xy}=\dfrac{\hbar}{i}
\sum_{\substack{ n k \sigma \\[1 pt] E_{n}<\epsilon_\mathrm{F}<E_{n+1}}}
\dfrac{\bra{n}\hat{j}_x\ket{n+1}\bra{n+1}\hat{j}_y\ket{n}-\bra{n+1}\hat{j}_x\ket{n}\bra{n}\hat{j}_y\ket{n+1}}
{(E_{n+1}-E_{n})^2}.
\label{eq_lsfin}
\end{equation} 
In this calculation, one has to evaluate the summation of matrix elements for current operator $\hat{j}$ such as $\bra{n}\hat{j}\ket{n+1}$ or $\bra{n+1}\hat{j}\ket{n}$, where $\ket{n}$ or $\ket{n+1}$ denote the states connected by this operator. 
In the present case of graphite in the quasi-quantum limit, these states should be chosen from the sub-bands having different Landau indices by number of 1, and same wave numbers and spins. Since the conductivity is determined by the summation of these matrix elements of the inter-sub-band transitions over a wide region of the reciprocal space, gap formation at a part of  the sub-band may cause the limited contribution to the Hall conductivity in the density wave state while the gap is considerably smaller than the energy difference in the adjacent sub-bands. 
In other words, comparison between the experimental data of $\sigma_{xy}$ and such calculations will provide in-depth evaluation for several models of sub-band structures above 53 T.
We leave it as an open problem because this calculation is beyond the scope of the present study.

\begin{figure}
\begin{center}
\includegraphics[width=7.5cm]{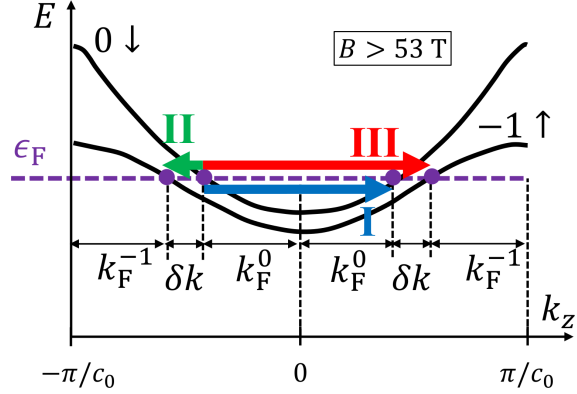}
\caption{
(Color online) The proposed sub-band structure realized above 53 T and possible three nesting vectors, labeled I, II, and III.
}
\label{ex}
\end{center}
\end{figure}
In the present study, we consider the possible sub-band structure through more qualitative argument. Firstly the observed small $\rho_{xy}$ seems to fit with the symmetric sub-band structure with respect to the Fermi level as illustrated in Fig. \ref{picandzz}(a) rather than the asymmetric one caused by depopulation of $-1$$\downarrow$ sub-band only.
Secondly, hole-doping enlarges the distance between the Fermi points in the $-1$$\downarrow$ sub-band as shown in Fig. \ref{picandzz}(b), which requires a large field to detouch from the Fermi level. On the contrary, our experimental results show smaller transition fields for $\alpha'$ in the neutron-irradiated graphite that has a larger hole concentration.
Thirdly the nearly simultaneous depopulation of two sub-bands seems to be natural in the point of view of the charge neutrality condition. Therefore, we interpret the behavior of Hall resistivity and magnetization as features of the realization of the quantum limit state in which only two sub-bands (electron-like $0$$\downarrow$ and hole-like $-1$$\uparrow$ sub-bands) cross the Fermi level. 

The observed enhancement of $\rho_{zz}$ above 53 T suggests the emergence of a novel density wave state between 53 T and 75 T in the quantum limit state.\cite{Fauque} In this state, the possible nesting vectors are shown in Fig. \ref{ex} by arrows labeled I, II, and III. Here, we suppose simultaneous nesting between the 
residual two Fermi points connected by  the same wave vector for the each state to reproduce the observed insulating behavior, although the nesting vectors are not described in Fig. \ref{ex} for clarity. The vector I represents the charge density wave phase, whereas the inter-sub-band nesting II and III denote the excitonic phase\cite{YF, manybody, Fenton, Jerome}.
According to the original argument of the excitonic insulator\cite{Jerome}, the present gapped state caused by the instability of the Fermi surface can be expressed by the BCS-like formulation, and called excitonic BCS-like state.
As stated in the Ref. \ref{ref_TT}, the charge density wave phases can be less stable because of the direct Coulomb interaction. Therefore, gap will be formed at the Fermi energy by inter-sub-band nesting with vector II or III, and simultaneously between the residual two Fermi points. 
Taking into account the charge neutrality condition ($k_\mathrm{F}^\mathrm{0}=k_\mathrm{F}^\mathrm{-1}$), the length of the vector III ($2k_\mathrm{F}^\mathrm{0}+\delta k$) is locked to $\pi/c_0$.
We suppose such a commensurate modulation can be favorable than the incommensurate one characterized by the vector II.
In this commensurate state, the carriers in a pair have finite average momentum, i. e. $(k_\mathrm{F}^\mathrm{0}+\delta k)-k_\mathrm{F}^\mathrm{0}=\delta k$, analogous to the FFLO phase in superconductors\cite{Fulde, Larkin}.
With increasing magnetic field, the value of $2k_\mathrm{F}^\mathrm{0}+\delta k$ remains $\pi/c_0$, whereas the $\delta k$ becomes larger.
The reported additional anomaly at 75 T might be a transition caused in the process of increasing the $\delta k$. The true nature of this transition will be clarified through additional experiments in the future.

\section{Conclusion}
The transport and magnetic properties of graphite were investigated in pulsed magnetic fields up to 75 T. The Hall resistivity becomes nearly zero as the applied magnetic field approaches 53 T. In addition, an anomaly is observed in differential magnetization at about 53 T. This transition field is reduced by doping holes. These results support the realization of the quantum limit state, where only two sub-bands (spin polarized one electron-like and one hole-like sub-band) cross the Fermi level, in the field region over 53 T. Formation of the density wave state in this situation can be regarded as the excitonic BCS-like state.
\begin{acknowledgment}
We thank Y. Iye for supplying various samples, and T. Osada and Y. Takada for valuable discussions and comments.
\end{acknowledgment}

\end{document}